\documentclass[journal=jacsat,manuscript=article]{achemso}

\mciteErrorOnUnknownfalse
\usepackage[UKenglish]{babel}
\usepackage{latexsym}
\usepackage{graphicx}	
\usepackage{amssymb}
\usepackage{amsmath}

\usepackage[T1]{fontenc} 

\author{Pedro Soubelet}
\affiliation{Walter Schottky Institut and Physik Department, Technische Universit\"{a}t M\"{u}nchen, Am Coulombwall 4, 85748, Garching, Germany.}
\author{Julian Klein}
\affiliation{Walter Schottky Institut and Physik Department, Technische Universit\"{a}t M\"{u}nchen, Am Coulombwall 4, 85748, Garching, Germany.}
\alsoaffiliation{Department of Materials Science and Engineering, Massachusetts Institute of Technology, Cambridge, Massachusetts 02139, USA}
\author{Jakob Wierzbowski}
\affiliation{Walter Schottky Institut and Physik Department, Technische Universit\"{a}t M\"{u}nchen, Am Coulombwall 4, 85748, Garching, Germany.}
\author{Riccardo Silvioli}
\affiliation{Walter Schottky Institut and Physik Department, Technische Universit\"{a}t M\"{u}nchen, Am Coulombwall 4, 85748, Garching, Germany.}
\author{Florian Sigger}
\affiliation{Walter Schottky Institut and Physik Department, Technische Universit\"{a}t M\"{u}nchen, Am Coulombwall 4, 85748, Garching, Germany.}
\author{Andreas~V. Stier}
\affiliation{Walter Schottky Institut and Physik Department, Technische Universit\"{a}t M\"{u}nchen, Am Coulombwall 4, 85748, Garching, Germany.}
\author{Katia Gallo}
\affiliation{Department of Applied Physics, KTH Royal Institute of Technology, SE-106 91 Stockholm, Sweden.}
\author{Jonathan~J. Finley}
\email{finley@wsi.tum.de}
\affiliation{Walter Schottky Institut and Physik Department, Technische Universit\"{a}t M\"{u}nchen, Am Coulombwall 4, 85748, Garching, Germany.}

\title{Charged exciton kinetics in monolayer MoSe$_2$ near ferroelectric domain walls in periodically poled LiNbO$_3$}


\begin{document}

\begin{abstract}
Monolayers of semiconducting transition metal dichalcogenides are a strongly emergent platform for exploring quantum phenomena in condensed matter, building novel opto-electronic devices with enhanced functionalities. Due to their atomic thickness, their excitonic optical response is highly sensitive to their dielectric environment. In this work, we explore the optical properties of monolayer thick MoSe$_2$ straddling domain wall boundaries in periodically poled LiNbO$_3$. Spatially-resolved photoluminescence experiments reveal spatial sorting of charge and photo-generated neutral and charged excitons across the boundary. Our results reveal evidence for extremely large in-plane electric fields of 3000\,kV/cm at the domain wall whose effect is manifested in exciton dissociation and routing of free charges and trions toward oppositely poled domains and a non-intuitive spatial intensity dependence. By modeling our result using drift-diffusion and continuity equations, we obtain excellent qualitative agreement with our observations and have explained the observed spatial luminescence modulation using realistic material parameters.
\end{abstract}


For integrated opto-electronic and quantum photonic devices, the ability to combine different materials having complementary functionalities is key to achieving high performance. Two-dimensional (2D) transition metal dichalcogenides (TMDs), having the chemical formula MX$_2$ (M=Mo,W and X=Se,S), are of strong current interest since they are direct gap semiconductors in the monolayer form, with bandgaps that are tunable throughout the visible spectral range. They host strongly bound excitons (Bohr radii $a_B \sim 1.1$-$1.8\,$nm, binding energy $E_b \sim 300$-500\,meV) \cite{goryca2019revealing(19)} that remain stable up to room temperature \cite{mak2010atomically}. Moreover, they can be readily van-der-Walls bonded onto a wide range of different substrates \cite{Wang-NatureNanotechnology7-699(12)}.  Unlike conventional semiconductors, the atomic thickness and 2D nature of TMDs is such that their properties change depending on the substrate onto which they are placed.  This opens up new routes to tailor the exciton energy landscape by intentionally placing TMDs onto substrates with spatially varying dielectric properties \cite{price2019engineering, wang2019inlaid, stanford2018emerging, wei2017size, gong2014vertical}. For example, lateral quantum confinement can enhance quasiparticle correlations, opening the way to study collective behaviour and many-body physics \cite{sigl2020condensation, ghiotto2020magic, regan2020mott, Raja2017, ugeda2014giant, xu2020creation, yu2017moire}. 

While dielectric engineering is sufficient for trapping neutral excitonic complexes, the manipulation of free-charges and charged excitons and the exciton ionization needed for efficient photodetection requires strong electric fields \cite{lau2018interface, 10.1126/science.1244358(13), 10.1038/s41565-017-0030-x(18), 10.1038/ncomms2498(13)}. Gate tunable devices with e.g. nanoscale metallic contacts can result in strain fields and large Schottky barrier heights. Moreover, proximity-induced electric fields arising from Fermi-level pinning produce band bending and local potential barrier in the TMD \cite{10.1038/s41565-017-0030-x(18), liu2016van}.  As such, new approaches are needed to produce high electric fields over lengthscales comparable to the exciton Bohr radius.  Lithium niobate is an exceptional material for integrated opto-electronics and photonics; it has a broad transparency window spanning the range 0.35-5\,$\mu$m, strong optical activity that can enable optical phase modulation as fast as 100\,GB/s \cite{guarino2007electro, wang2018integrated, he2019high} and can be integrated on oxide sacrificial layers using CMOS compatible processes to produce low loss waveguides ($< 2.7\,$dB/m \cite{zhang2017monolithic}). Moreover, periodically poled lithium niobate (PPLN) is ferroelectric \cite{10.1021/acsnano.8b09800(19), 10.1021/acsomega.6b00302(16), li2020polar} and has extremely large in-plane surface charge densities within individual domains \cite{10.1038/ncomms15768(17), PhysRevB.89.226101}.  Atomically sharp ferroelectric interfaces (domain walls) in combination with the insulating nature of PPLN, result in extremely intense in-plane electric fields \cite{morozovska2008effect}, capable of driving e.g. localized photochemical reactions \cite{carville2012photoreduction, hanson2006fabrication} and ferroelectric lithography of multicomponent nanostructures \cite{kalinin2004ferroelectric}. Furthermore, the size and shape of such domains can be freely tailored during the ferroelectric pattering \cite{10.1021/acsnano.8b09800(19), 10.1021/acsomega.6b00302(16), li2020polar, gallo2006bidimensional}. Placing TMD monolayers and heterostructures onto ferroelectric materials provides the possibility to locally tune optical \cite{10.1021/acsnano.8b09800(19), 10.1021/acsomega.6b00302(16), li2020polar, lv2019reconfigurable} and electronic properties \cite{li2020polar, xiao2017ferroelectric, lipatov2019nanodomain}, engineer the local charge density landscape \cite{10.1021/acsnano.8b09800(19)} and facilitate study of fundamental properties such as in-plane exciton polarizability and dissociation \cite{kamban2020interlayer, PhysRevB.94.245434}.

Here, we explore the excitonic photo-physics of monolayer MoSe$_2$ exfoliated onto a PPLN crystal. By performing low-temperature ($T=17$\,K) micro-photoluminescence ($\mu$PL) measurements, we observe separation and localization effects of charged excitons and free carriers in the flake close to the domain wall (DW) between positively ($P^+$) and negatively ($P^-$) poled ferroelectric domains, orientated perpendicular to the PPLN surface.  Due to the very large in-plane electric fields that form at the ferroelectric DW boundaries \cite{morozovska2008effect}, free-charges and photogenerated charged excitons respond to the DW by forming a characteristic charge and photo-generated texture. In addition, we observe that the DW induces strong modulation of the relative energies of the neutral and charged excitons in the monolayer MoSe$_2$. We theoretically model our observations using drift-diffusion and continuity equations \cite{jimenez2012drift, Ivanov_2002} and find excellent qualitative agreement with our experimental results. Our simulations allow us to evaluate the dynamics of charged carriers in the system and show that the combined structure of DW and MoSe$_2$ behaves like a nanometer scale lateral p-n homojunction. These results are an important step towards integrated optoelectronic devices utilizing atomically thin semiconductors on PPLN and provide new insights into the excitonic photophysics subject to very large electric fields.

\begin{figure}[t!!]
\includegraphics*[keepaspectratio=true, clip=true, angle=0, width=.7\columnwidth, trim={4mm, 2mm, 3mm, 2mm}]{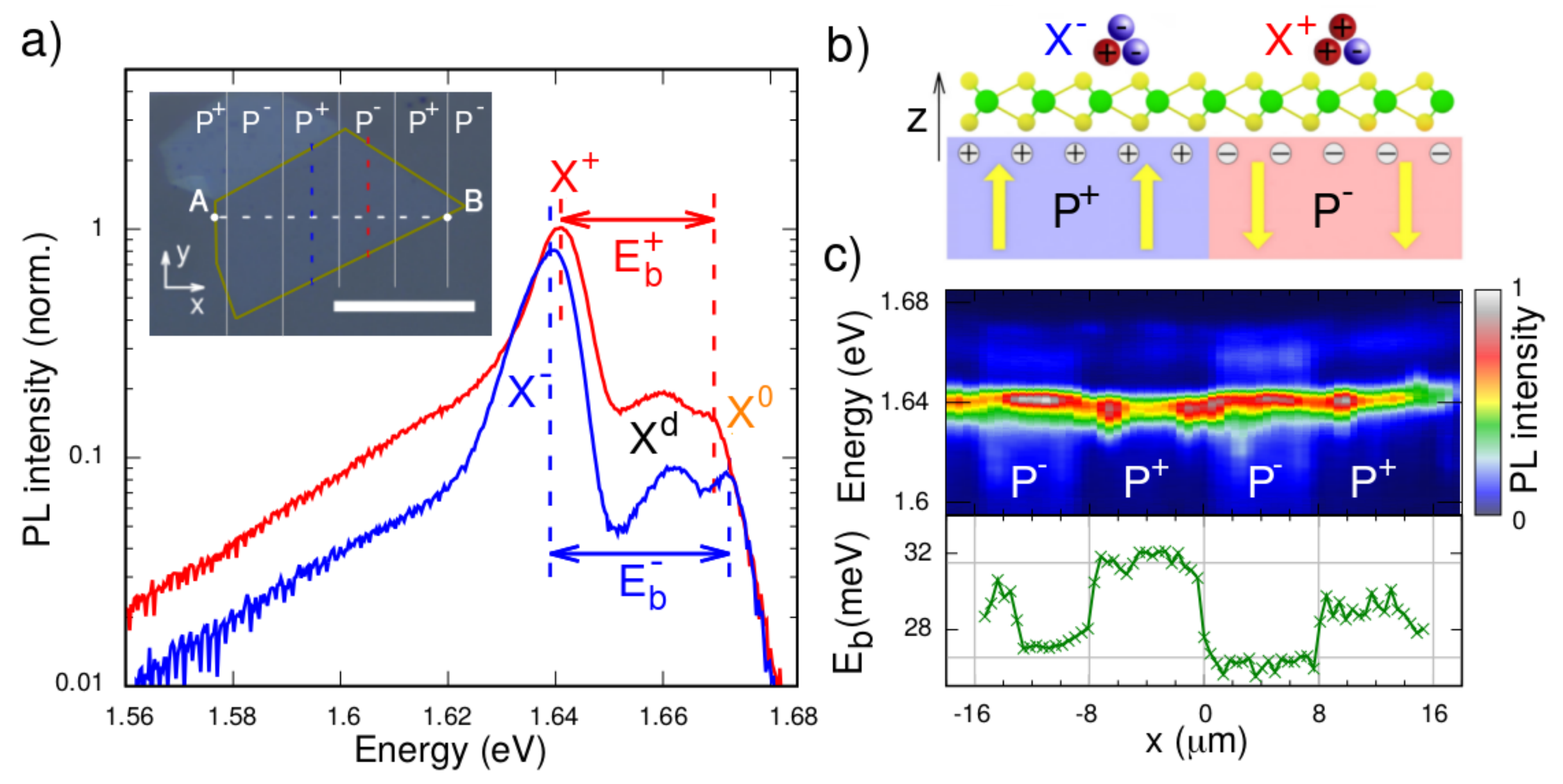}
\caption{\textbf{Monolayer MoSe$_2$/PPLN device and $\mu$PL experiments. a)} Spatially averaged low-temperature $\mu$PL spectra recorded at the center of each domain, \textit{$P^+$} and \textit{$P^-$} in blue and red, respectively. The spectra show the trion emission ($X^{+/-}$), the neutral exciton emission ($X^0$) and the emission from excitons trapped at defect ($X^d$). Inset: optical micrograph of the sample: yellow lines indicate the contour of the MoSe$_2$ and the grey lines mark the PPLN DWs separated by $\sim8\,\mu$m. The red and blue dotted lines indicate the transects along which the spectra were recorded. The scale bar is $20\,\mu$m. \textbf{b)} Schematic illustration of the sample with positive and negative ferroelectric polarity and its induced surface charge. Consequently, negatively (positively) charged excitons ($X^{\pm}$) are expected to form in the center of the $P^+$ ($P^-$) domains. \textbf{d)} Top panel: PL false colour map of the sample along the line denoted A-B in the inset of a). The map shows the modulation in energy and emission intensity across the different domains. Bottom panel: Modulation of the trion binding energy ($E_b$) across the different domains.}
\label{figura:1}
\end{figure}

\section{Results and Discussion}

We recorded $\mu$PL data with an optical CW power of $P \sim 20\,\mu$W ($E=1.96\,$eV) focussed to a diffraction limited spot (100x objective, NA=0.7). Figure \ref{figura:1}a shows a typical spectrum obtained well within $P^+$ (blue) and $P^-$ (red) domains. The inset of figure \ref{figura:1}a shows an optical micrograph recorded from the studied sample. The picture shows the monolayer thick MoSe$_2$ crystal outlined by the yellow line and the DWs are denoted by grey vertical lines. The sample is exfoliated onto the PPLN substrate such that the monolayer spans multiple domains. For details of how we determined  the  PPLN  domains, see the Additional Information \cite{SOM}. The spectra presented are the spatially averaged spectra along vertical blue and red transects illustrated by the dotted lines in the inset. Within both domains we observe $X^0$ emission close to $\sim 1.67\,$eV, consistent with other reports \cite{cadiz2017excitonic(17), 10.1038/s41598-017-09739-4(17)}. However, in both domains the dominant emission feature arises from trions; around $\sim 1.64$\,eV \cite{cadiz2017excitonic(17), 10.1038/s41598-017-09739-4(17)}. In addition, we observe a luminescence peak at $\sim1.66\,$eV, labelled $X^d$ on Fig. \ref{figura:1}b, which does not occur when the same crystal is exfoliated onto SiO$_2$ or hexagonal boron nitride. We identify this peak as arising from defect induced transitions \cite{hong2015exploring(15)} and do not discuss it for the remainder of this manuscript. The electric polarization of the LN ($P^{+,-}$) in a direction normal to the plane of the MoSe$_2$ monolayer ($z$-axis) results in a large density of fixed surface polarization charges \cite{10.1021/acsomega.6b00302(16)}, as indicated in the schematic of Fig. \ref{figura:1}b. These surface charges attract oppositely charged mobile carriers from the native doping of the monolayer and thus, at the center of each domain, a neutral exciton ($X^0$) can bind an available additional electron or hole to form positive ($X^+$) or negative ($X^-$) trions, respectively. The pronounced emission of trions is thus an evidence of a large accumulation of holes and electrons in the $P^+$ and $P^-$ and therefore, the PPLN creates an electronic landscape that is similar to a p-n junction. 

\begin{figure}[t!!]
\includegraphics*[keepaspectratio=true, clip=true, angle=0, width=.7\columnwidth, trim={3mm, 6mm, 4mm, 4mm}]{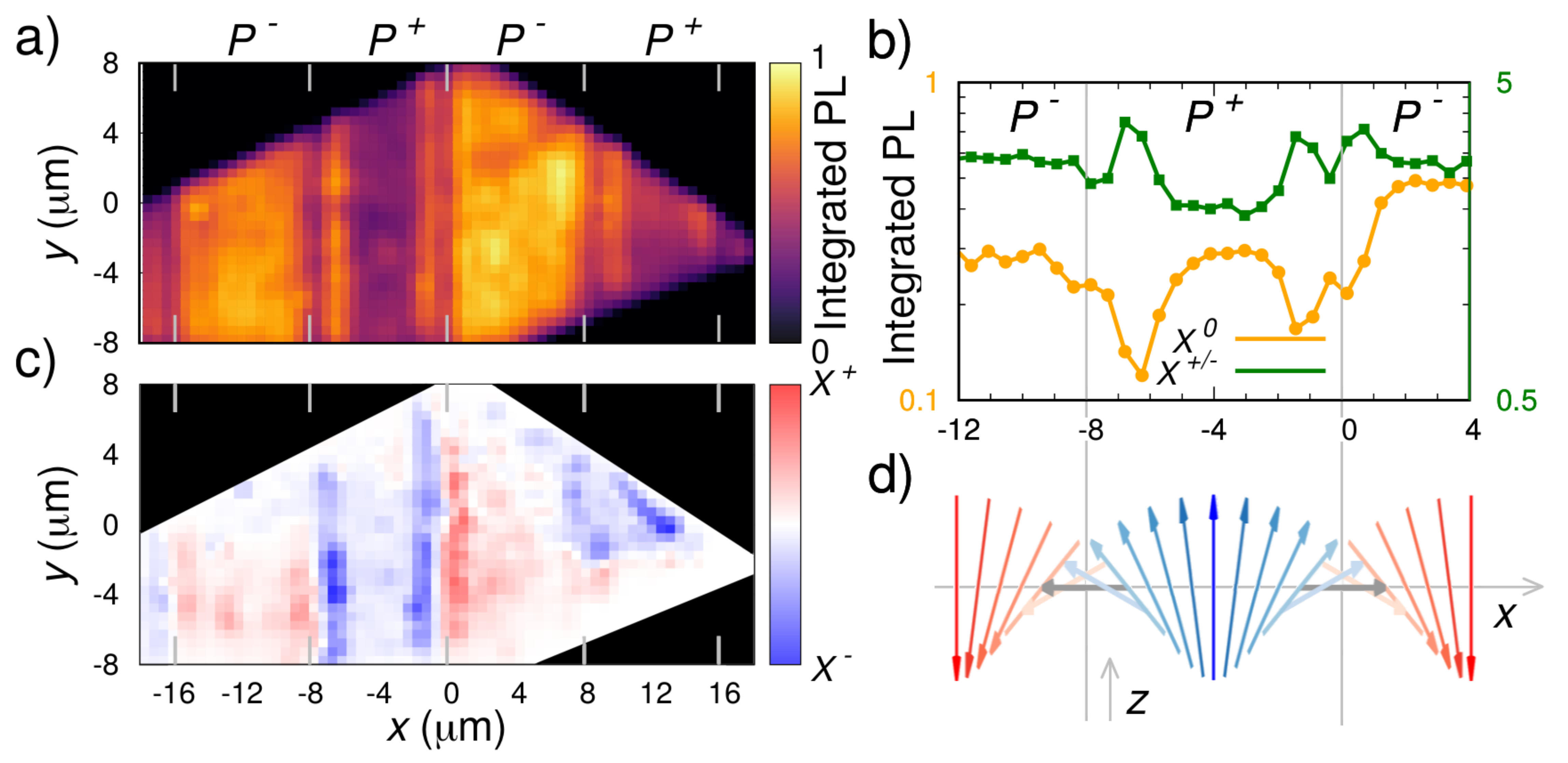}
\caption{\textbf{Spatially dependent $\mu$PL data recorded across the PPLN domains. a)} Integrated PL map of the MoSe$_2$ revealing the spatial modulation of the overall intensity and spectral of the PL emission. The scale bar is 5\,$\mu$m. \textbf{b)} Spectrally integrated PL intensity along the $x$-direction of the charged ($X^{+/-}$) and neutral ($X^0$) excitons. Around the DW, trion emission is enhanced while neutral exciton emission decreases. \textbf{c)} Trion intensity map, considering that $E_b(X^-)\sim 31$\,meV and $E_b(X^+)\sim 26$\,meV (Fig. \ref{figura:1}d), trion PL intensities centered at energies below (above) 28.5\,meV are presented in red (blue) scale. The scale bar is 5\,$\mu$m. \textbf{d)} Schematic representation of the electric field at the interface between the regions \textit{$P^+$} and \textit{$P^-$}.}
\label{figura:2}
\end{figure}

We continue to explore the behavior of the steady state PL when spanning different domains in the MoSe$_2$. Fig. \ref{figura:1}c shows a series of spatially dependent PL spectra presented as a false color representation along the axis A-B, denoted by the white dotted line in the inset of Fig. \ref{figura:1}a. The domains are clearly visible due to an enhanced background PL emission in the $P^-$ regions, possibly due to negatively charged chalcogen vacancies in the material \cite{10.1038/ncomms2498(13), 10.1021/acsnano.8b09800(19), 10.1038/s41598-017-09739-4(17), 10.1103/PhysRevLett.109.035503(12), hong2015exploring(15), shepard2017trion(17), sercombe2013optical}. Most prominently, the energy of both $X^+$ and $X^-$ trion peaks ($\sim1.64\,$eV) shifts abruptly across the interface, while simultaneously, the energy of the neutral exciton shifts in the opposite direction to the trion. Hence, the binding energy of the trion ($E_b = E_{X^0} - E_{X^{+/-}}$ \cite{10.1038/ncomms2498(13)}) changes abruptly across the DW interface. We quantify the change of $E_b$ by fitting the PL spectra and statistically analyzing the relative peak positions across the domains. The results of this analysis are plotted in the bottom panel of Fig. \ref{figura:1}c, which shows a clear modulation of $E_b$ induced by the different polarization states in the PPLN. As the binding energy of a trion is proportional to the effective masses of the individual particles of the bound state (e.g. $m_{X{^-}} = 2\cdot m_e + m_h$, where $m_e$ is the electron mass and $m_h$ is the hole mass) \cite{10.1038/ncomms2498(13), xiao2012coupled(12), xenogiannopoulou2015high, mo2016spin, zhang2014direct}, we attribute the variation in $E_b$ to the different charge of the trions on each side of the interface. Consistent with literature \cite{10.1021/acsnano.8b09800(19)}, we find that the $P^+$ domains feature $X^-$, while in the $P^-$ domains we predominantly observe $X^+$ emission. The modulation in $E_b$ occurs abruptly within the step size of our lateral scanning stage, $\Delta x = 250\,$nm. Such abrupt modulations of the neutral exciton binding energy have been realized in monolayer TMDs by encapsulation \cite{Raja2017}, however here we do not modulate the dielectric environment of the system, but rather spatially modulate the local charge density in the TMD monolayer via the PPLN surface charge. The sharp modulation in $E_b$ is an indication that the electric field gradient across the DW is very large, as discussed in detail below.

We now focus our analysis on the effect of the PL intensity close to the DW. Even though the modulation of the spectrally integrated PL intensity and $E_b$ is a direct consequence of the underlying domain structure, they show different behaviour. As can already be seen in the raw spectra plotted in Fig. \ref{figura:1}c, the PL intensity is modulated on both sides of the interface, while the modulation in $E_b$ follows a step-like pattern. Figure \ref{figura:2}a presents a map of the total integrated PL across the sample. It shows that the PL intensity reflects the geometry of the underlying DW boundaries in the $x$-direction, but also that satellite stripes of intensity are observed orientated along the $y$-direction, positioned $x\sim 1-2\,\mu$m away from the DW boundary.

To elucidate this unexpected behavior, we plot in Fig. \ref{figura:2}b the averaged total intensity integrated over the spectral range of the \textit{$X^{+/-}$} and $X^0$ emission as a function of the $x$-position of the laser spot.  On either side of the DW boundary, we observe that the charged exciton emission increases while simultaneously, the neutral exciton emission decreases in an anticorrelated manner. This behavior in combination with the step like $E_b$ modulation is consistent with an increase in the electron and hole density on each side of the interface that facilitates the formation of positive and negative trions and reduces the local formation probability for $X^0$. Thereby, the electric field across the ferroelectric DW, which has a thickness of few nanometers \cite{liu2014theoretical(14)}, promotes the dissociation of excitons and the separation of trions in a micrometer scale according to their charge species.

\begin{figure}[ttt!!]
\includegraphics*[keepaspectratio=true, clip=true, angle=0, width=.7\columnwidth, trim={2mm, 3mm, 0mm, 0mm}]{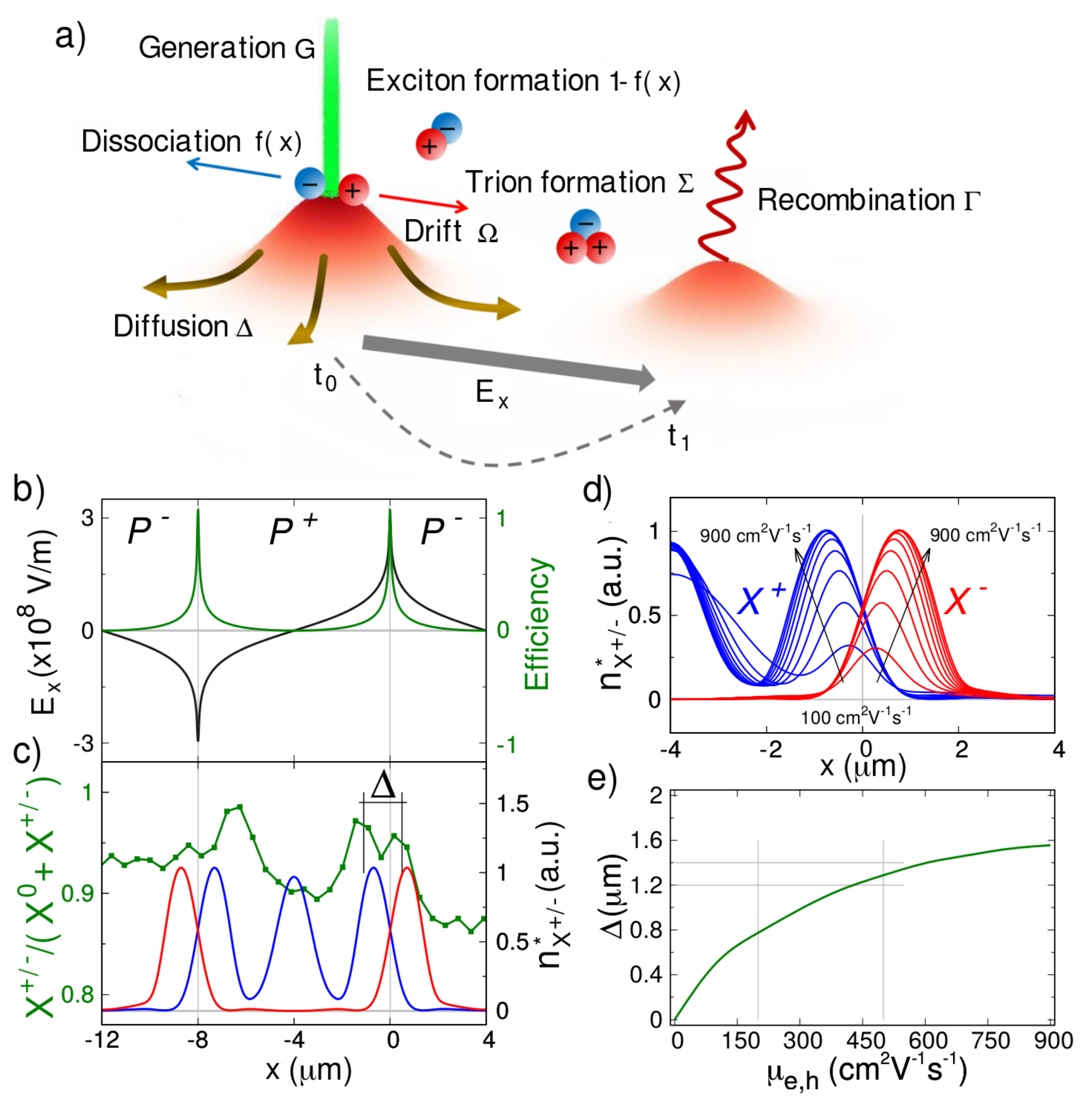}
\caption{\textbf{Drift and diffusion continuity rate equation model. a)} Schematic representation of the drift-diffusion model used to simulate our experiment . The laser generates electrons and holes in the sample with a rate $G$ withing a Gaussian laser spot focus. These charges have a probability $f(x)$ to drift within the local in-plane electric field and a probability $1-f(x)$ to bind together to form an exciton. While the diffusion $\Delta$ affects all the particles, the in-plane electric field $E_x$ pushes the charged particles in opposite directions (drift $\Omega$) and localizes them on opposite sides of the DW. Eventually, a neutral exciton has a probability $\Sigma$ to form a trion by the capture of an electron or a hole. Finally, the excitons recombine, and the luminescence is localized mainly on both sides of the DW. \textbf{b)} Calculated in-plane electric field for a positive/negative charge density uniformly distributed in the $P^+$/$P^-$ region (black) and electron-hole generation efficiency $f(x)$ (green) proportional to $E^2_x$. \textbf{c)} \textit{$X^{+/-}/(X^{+/-}+X^{0})$} determined with the fitted PL intensities and the calculation of the positive and negative trion densities across the domains. The spatial distance between the positive and negative trion peaks ($\Delta$) is marked on the experimental data. \textbf{d)} Negative (blue) and positive (red) trion density calculation for different electron/hole mobility from 100 to 900\,cm$^2$V$^{-1}$s$^{-1}$. \textbf{e)} Spatial distance between the positive and negative trion peaks in d). The experimental $\Delta$ (horizontal lines) and the electron mobility (vertical lines) from ref. \cite{chamlagain2014mobility} (500\,cm$^2$V$^{-1}$s$^{-1}$) and \cite{larentis2012field} (200\,cm$^2$V$^{-1}$s$^{-1}$) are depicted in grey. The latter correspond to higher temperatures than that of our experiments ($\sim80\,$K).}
\label{figura:3}
\end{figure}

The separation of positive and negative trions along the macroscopic PPLN DW is presented in Fig. \ref{figura:2}c, and is one of the main experimental findings of this letter. In this map, we colour code the trion intensity based on the fact that $E_b(X^-)\sim 31$\,meV and $E_b(X^+)\sim 26$\,meV (Fig. \ref{figura:2}d). Trion intensities at energies below (above) 28.5\,meV are presented in red (blue) scale to highlight their different charge species. In the center of each ferroelectric domain, the induced electric field is oriented in the direction of its polarization ($z$-direction). However, the separation of charged excitons in a direction defined by the sign of their charge is a clear fingerprint of an in-plane electric field perpendicular to the interface, along the $x$-direction. Such an electric field is provided by the formation of a N\'{e}el-type DW \cite{liu2014theoretical(14), 10.1038/ncomms15768(17)} as depicted in Fig. \ref{figura:2}d. As consequence, optically-generated electron-hole pairs are dissociated and the resulting free carriers are efficiently separated by the DW. In our experiment, we scan the diffraction limited laser focal volume at a position $x_0$ across the sample and detect the PL from the same spot. Nevertheless, due to the strong in-plane electric fields, specifically at the DW, the exciton generation as well as the drift and diffusion of charged excitons and excess carriers is strongly dependent on $x_0$. The complete process is illustrated schematically in figure \ref{figura:3}a.

By solving a closed set of coupled differential equations that include the optical generation, exciton and trion formation, drift and diffusion, we qualitatively describe the time evolution of the different particle densities until they reach the steady state. Details of our approach can be found in the Additional Information \cite{SOM}. Typical results are presented in figure \ref{figura:3}. Figure \ref{figura:3}b shows the spatial in-plane electric field profile experienced by the MoSe$_2$ monolayer at the surface of the PPLN substrate. In this calculation, we use a uniform fixed positive and negative charge density of $\pm2\,\mu$C$/\mu$m$^2$ in the $P^{\pm}$ domains, in accordance with the charge density deduced from the modulation of the refractive index of the material, as shown in the Additional Information \cite{SOM}. The calculated electric field strength at the interface is extremely high, of 3000\,kV/cm and comparable to the breakdown field of hexagonal boron nitride \cite{lee2011electron}. This opens the possibility to perform the experiments shown here in a gate-tunable device made from stacked van der Waals materials. As the energy of free carriers exposed to the electric field has to overcome the large exciton binding energy in the monolayer MoSe$_2$, we model the probability of exciton dissociation, $f(x)\propto E^2_x$ (see Fig. \ref{figura:3}b), \cite{goryca2019revealing(19)}.

With the in-plane electric field and the generation efficiency as input in our coupled rate equations (Figs. \ref{figura:3}b), we calculate the steady state of the different particle distribution and obtain, in this way, their spatial density profile across the PPLN domains as the laser spot is scanned across the sample. The spatial redistribution of luminescence intensity along the PPLN domain edges is due to the spatial density profile of excitons and trions. Since the equations are highly interdependent and sensitive to material properties, here we only focus on the dependence on carrier mobility, while additional more detailed information is presented in the Additional Information \cite{SOM}.  

Figure \ref{figura:3}c presents the calculated spatially dependent density of $X^{\pm}$, $n^*_{X^{\pm}}$, and compares it to the relative PL intensity profile of excitons and trions. The symbol $*$ denotes that we plot the observed density as the laser spot is scanned across the sample (see ref. \cite{SOM}). The simulation clearly shows that we can quantitatively describe the separation of charged excitons at the interface as observed in the experimental data of Fig. \ref{figura:2}. Positive excitons form predominantly on the $P^-$ side at the interface, but not at the DW boundary. In comparison, negatively charged trions form in the opposite side of the DW (see Fig. \ref{figura:2}c). The increase of the $X^-$ intensity towards the center of the $P^+$ domain is controlled by the native electron doping of MoSe$_2$, described above. Since the charge mobility depends on the effective mass and our results for the $X^+$ and $X^-$ binding energy supports the conclusion of similar effective electron and hole masses, we use an equal carrier mobility for both particles. In figure \ref{figura:3}d, we show the spatial distribution of $X^{\pm}$ density in the vicinity of the interface for a range of carrier mobilities. For a given electric field, free charges and trions ($e$, $h$ and $X^{\pm}$) are accelerated large distances away from the DW, an effect that defines the spatial position of the peak of the calculated $n^*_{X^{\pm}}$ distribution at the DW. The highly nonlinear amplitude of $n^*_{X^{\pm}}$ with excitation power is the result of a delicate interplay of electron/hole generation, drift, diffusion and exciton formation probability close to the interface (see Additional Information \cite{SOM}). We deduce and plot the separation of the $X^{\pm}$-luminescence peak ($\Delta$) in Fig. \ref{figura:3}e and find that the experimentally measured peak separation agrees well with our calculated values using a carrier mobility of $\mu = 500$\,cm$^2$V$^{-1}$s$^{-1}$. This value is in very good accord to previously published values for MoSe$_2$ at low temperatures \cite{chamlagain2014mobility}.  Remarkably, despite its simplicity our model qualitatively describes the relevant physics in this system, although we reiterate that the formation and recombination of excitons is a highly non-linear process that depends on several interralated material specific parameters.

\begin{figure}[ttt!!]
\includegraphics*[keepaspectratio=true, clip=true, angle=0, width=.7\columnwidth, trim={0mm, 0mm, 0mm, 0mm}]{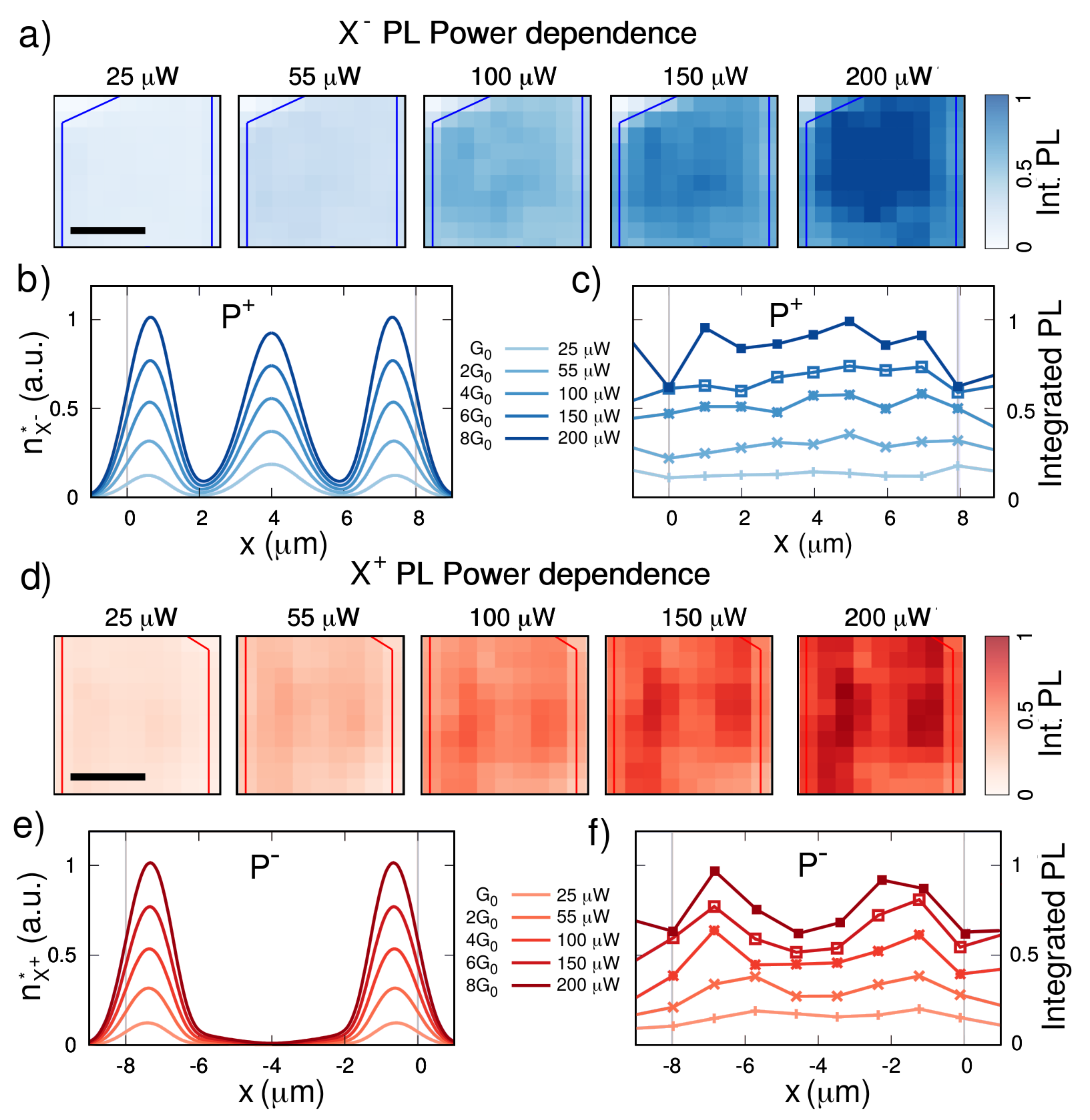}
\caption{\textbf{Spatially dependent $\mu$PL data recorded at different PPLN domains for increasing excitation power. a)} Integrated PL map of the MoSe$_2$ in the spectral region of the negative trion for different laser power intensities from 25\,$\mu$W to 200\,$\mu$W that shows a continuous growth of the PL in the $P^+$ domain. The scale bar is 4\,$\mu m$ \textbf{b)} Negative trion density calculation that shows a continuous growth by increasing the generation intensity ($G_0$). \textbf{c)} Spectrally integrated experimental $X^-$ PL intensity along the $x$-direction that shows a uniform intensity proportional to the excitation power. \textbf{d)} Integrated PL map of the MoSe$_2$ in the spectral region of the positive trion for different laser power intensities. The maps show a notorious increment in the PL signal close to the $P^-$ DW. The scale bar is 4\,$\mu m$. \textbf{e)} Positive trion density calculation for different generation intensities ($G_0$). \textbf{f)} Spectrally integrated experimental $X^+$ PL intensity along the $x$-direction.}
\label{figura:4}
\end{figure}

To further explore the consistency of the implemented model, we evaluate the PL dependence with the laser power from $25$ to $200\,\mu$W. Over the studied range, the intensity of $X^0$ shows a monotonic increase and therefore confirms that we probe the sample in a regime where saturation effects due to exciton-exciton annihilation are not significant. Figure \ref{figura:4}a shows a sequence of color maps of the integrated $X^-$ PL intensity recorded for increasing excitation levels. In the $P^+$ domain, the $X^-$ PL intensity is spatially homogeneous along the $x$-direction. Variations in the total intensity along the $y$-direction may be related to varying contact of the MoSe$_2$ to the substrate and edge effects. Figures \ref{figura:4}b and c present the modeled $X^-$ density and the experimental $X^-$ PL intensity along the $x$-direction for increasing generation rate of excitons $G$ and laser power, respectively. Once again, we find a linear dependence with increasing optical power, in a similar fashion as before.

As indicated above, the calculated and measured luminescence intensity across the domain is distributed close to the DW but also in the center due to the native electron density in the MoSe$_2$ layer. Although the calculated $X^-$ density is not uniform across the $x$-direction, it nicely reproduces the presence of negative trions in the center of the domain with an intensity that, in this regime, is also directly proportional to the generation rate. The lack of negative trion stripes close to the DW is most likely due to the limited spatial resolution of $\sim 1\,\mu$m in our experiment. Figures \ref{figura:4}d-f show the analogous dependencies of the integrated PL intensity for the $X^+$ on power and position. The significant difference with the $X^-$ case is the obvious spatial variation of the intensity across the domain, due to the concentration of the $X^+$ close to the interface and the lack of native charge doping in the center of the domain. Moreover, we reiterate that our simple calculations capture the salient photophysics in our experiments. The model can be improved to further enhance the agreement with the experimental results by introducing, for instance, electron-electron interactions that would take into account the electric field screening \cite{zipfel2020exciton, wagner2020propagation}, providing a more uniform and intense negative trion density in the center of the positive domains. Such approaches are, however, beyond the scope of our model. 

In summary, we observe clear evidence for the impact of surface polarization charges and local electric field gradients on the excitonic photophysics of monolayer MoSe$_2$ placed on PPLN. The ferroelectric domains separate and localize charged excitons according to their charge species. The impact of the surface polarization charges due to the PPLN is observed in spatially resolved $\mu$PL experiments by analysing the spectral position of exciton and trion emission. We modelled our findings using a self-consistent generation, drift-diffusion finite element model and obtained results that were in excellent qualitative agreement with our experiments. Our model is robust and can qualitatively describe the physics of charged exciton formation and spatial distribution by strong in-plane electric fields in the linear regime where exciton-exciton and electron-electron interactions are not relevant. The huge magnitude of the electric field set by the atomically flat PPLN substrate (3000\,kV/cm) is difficult to reach using other techniques, such as gated devices since it can easily cause the dielectric breakdown of the system. On the other hand, even when this issue may be solved by means of hBN encapsulation, the metallic contacts generate a depletion region in the electronic bands of the TMD that hinder the localization of the charged particles in small regions \cite{liu2016van, 10.1038/s41565-017-0030-x(18)}. Under these considerations, PPLN represents a highly interesting platform to manipulate and localize charges in TMDs, and paves the way towards integrated photonic structures and waveguides that combines LiNbO$_3$ and 2D materials.

\section{Methods} \label{Met}
\textbf{Sample fabrication.} The ferroelectric domain arrays were fabricated by bulk electric-field poling of 500\,$\mu$m-thick congruent $z$-cut LiNbO$_3$ \cite{gallo2006bidimensional}. The poling pattern was defined by means of periodic openings in a micrometer-thick photoresist mask (S1818) deposited on LiNbO$_3$ monodomain samples. The PPLN pattern consisted of 15 uniform bands of constant period ($\Lambda$), ranging from 14.5 to 16.5\,$\mu$m, in steps of 0.5\,$\mu$m. Local domain switching was performed at room temperature, by applying $\sim$60\,ms pulsed electric fields exceeding the LiNbO$_3$ coercive value (21\,kV/mm), with a uniform electrical contact on the $+z$ face and periodic liquid electrode contacts through the openings of the photoresist on the $-z$ face. After poling, the poling mask was removed by ultrasonic rinsing the samples in deionized water, acetone and isopropanol and N2 blow-drying. The resulting 1D ferroelectric arrays exhibited a periodic pattern with domains of alternating polarity ($\pm z$) along the crystallographic $x$-axis and with atomically-thin 250\,$\mu$m-long DWs aligned to the $y$-axis. 
Finally, monolayer MoSe$_2$ was transferred onto the PPLN substrate by the dry viscoelastic transfer method \cite{castellanos2014deterministic} such that the monolayer covered several PPLN domains. 

\textbf{$\mu$PL experiments.} All the photoluminiscence experiments where performed using a confocal microscope foccused to the diffraction limited spot of 0.7\,$\mu$m on the sample with a photon energy of 1.96\,eV ($\lambda=632$\,nm, HeNe laser). During the experiments the sample was kept at $T=17$\,K and the continuous wave excitation power was 20\,$\mu$W unless otherwise noted. For the statistical analysis of the PL spectra, we omitted data recorded within $1\mu$m distance to flake edges to avoid well known edge effects (defects, strain, etc.).

\begin{acknowledgement}

We gratefully acknowledge the German Science Foundation (DFG) for financial support via the clusters of excellence e-conversion (EXC 2089) and the Munich Center for Quantum Science and Technology (MCQST, EXC 2111).  Moreover, we gratefully acknowledge the DFG Priority Programme (SPP 2244) via FI 947/7-1, the BMBF project QuaDiQua and the Alexander von Humboldt Stiftung for support. K.G. gratefully acknowledges support from the Wallenberg Centre for Quantum Technology and the Swedish Research Council (VR) through grants no. 2018-04487 and 2016-06122.

\end{acknowledgement}

\subsection*{Competing financial interests} 
The authors declare no competing financial interest.


\section{Additional Information}

\subsection{Determination of the PPLN domains and in-plane electric field estimation}\label{1}

\begin{figure}
\includegraphics*[keepaspectratio=true, clip=true, angle=0, width=.75\columnwidth, trim={0mm, 0mm, 0mm, 0mm}]{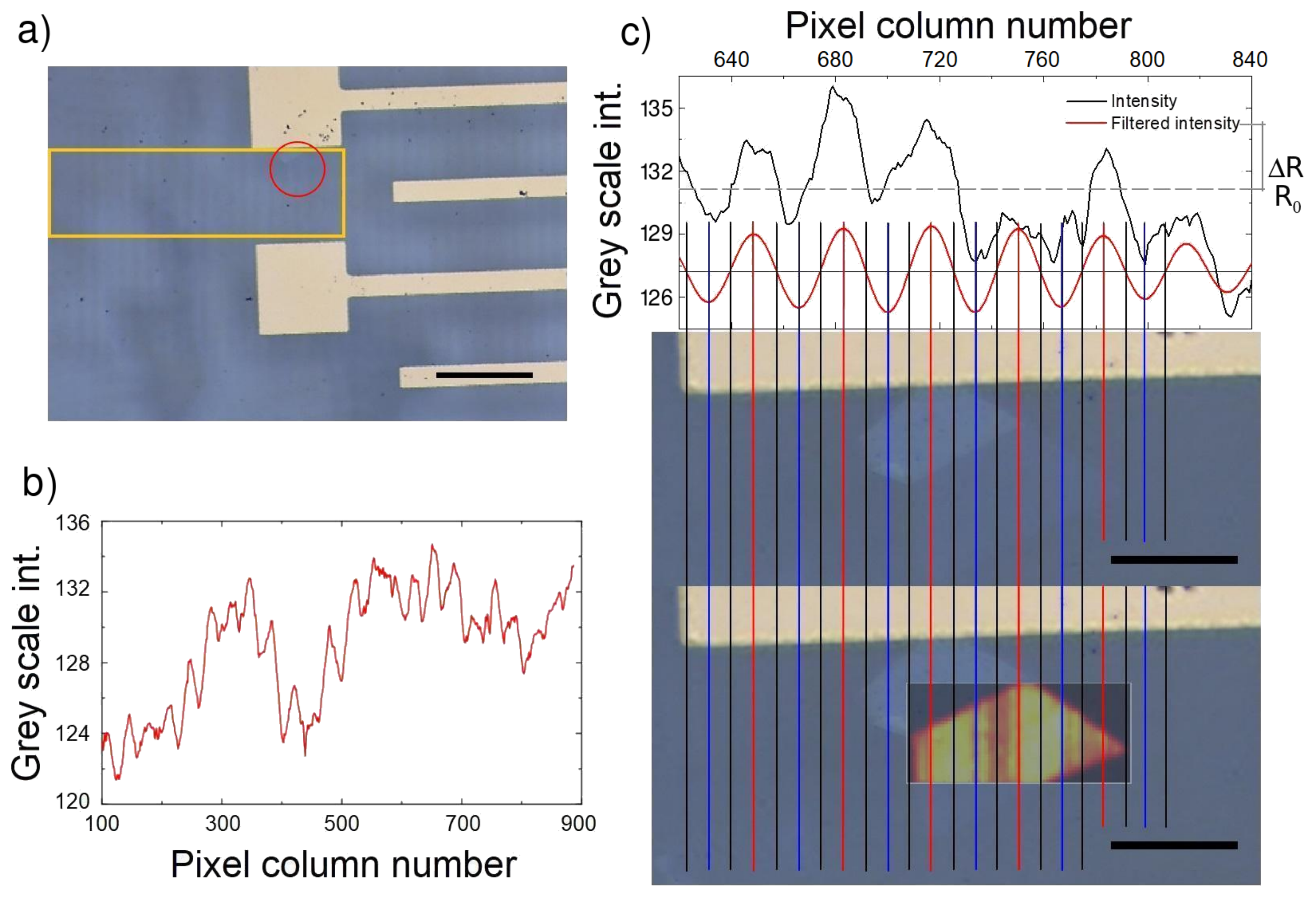}
\caption{\textbf{Determination of the PPLN edges. a)} Optical micrograph of the sample taken with a 20$\times$ objective. Evenly distributed in the LN substrate are vertical dark and bright stripes that correspond to variations in the reflectivity due to the PPLN domains. The red circle indicated the region in which the 1L-MoSe$_2$ is located and the yellow rectangle limits the region used to perform the image processing by averaging the pixel grey scale intensity in vertical direction. The scale bar correspond to 100\,$\mu$m. \textbf{b)} Grey scale intensity as function of the pixel column in the yellow region marked in a). \textbf{c)} Top panel shows a section of figure b) and the same signal band pass filtered at the frequency of the dark/bright stripes in the region of the 1L-MoSe$_2$ flake. The blue, black and red vertical lines mark the positions of the minima, zeros and maxima of the filtered signal, respectively. These vertical lines are extended over the optical images in the lower panels. The central panel shows an optical micrograph of the sample taken with a 20$\times$ objective overlapped with an optical micrograph taken with a 100$\times$ objective. The lower panel overlaps the integrated PL intensity map of the sample. The scale bar correspond to 20\,$\mu$m.}
\label{figura:SM1}
\end{figure}

Figure \ref{figura:SM1}a presents an optical micrograph of the sample taken with a 20$\times$ objective. The light coloured structure correspond to metallic contacts used to ground the sample and avoid its electrostatic doping. In this figure, vertical dark and bright stripes can be clearly identified, corresponding to intensity variations of the reflected light in the substrate due to different ferroelectric domains. The red circle in the figure indicates the region in which the monolayer MoSe$_2$ flake is located and the yellow rectangle limits the region selected to post process the picture by vertically averaging the pixel gray scale intensity. As the horizontal intensity alternation pattern is not observed in optical micrographs taken with higher magnification objectives, in order to identify the DW in the monolayer it was necessary to process this optical image and combine it with optical images taken with higher magnification objectives. The pixel intensity variations are presented in Fig. \ref{figura:SM1}b. The upper panel of figure \ref{figura:SM1}c shows an enlargement of the curve in b in the region of the monolayer, and the band-pass filtered data. The central panel presents the overlap of optical images taken with 20$\times$ 100$\times$ objectives and the lower panel overlaps to this picture the integrated total PL intensity map. The blue, black and red vertical lines indicate the position of minima, zeros and maxima of the filtered signal in the upper panel, respectively. By direct inspection between the different regions in the PL map with the maxima and minima of the grey scale curve (red and blue vertical lines), it is possible to find that the pixel intensity oscillations are $90^{\circ}$ out of phase with respect to the different PL intensity regions. In other words, the extremes of the reflectivity curve coincide with the DWs and not with the center of each ferroelectric domain. This observation is consistent with the existence of an in-built electrostatic field at the DW, that modulates the refractive index of the PPLN due to Pockels effect \cite{boyd2019nonlinear}.

The refractive index modulation amplitude $\Delta n$ can be directly inferred from the variation $\Delta I$ in the grey scale intensity retrieved from the optical micrograph. Specifically, assuming that the grey scale pixel intensity $I(x)$ is proportional to the reflectivity of the PPLN ($R(x)$) and applying the Fresnel equations, we get
\begin{equation}
R(x) = I(x)K = \left(\frac{n(x)-1}{n(x)+1}\right)^2,
\label{eq1}
\end{equation}
where $K$ is the proportionality constant and $n(x)$ is the spatially-varying refractive index, oscillating around the unperturbed value, corresponding to the ordinary refractive index of LN,  $n_0=2.176$, with an amplitude $\Delta n$ due to the Pockels effect perturbation. A first order expansion of this expression provides the relationship between $\Delta I$ and $\Delta n$:
\begin{equation}
K\frac{(n_0+1)^3}{4(n_0-1)} \Delta I = \Delta n,
\label{eq2}
\end{equation}
From figure \ref{figura:SM1}c and equation \ref{eq1} it is possible to estimate $I\simeq 131$,  $\Delta I \simeq 4$, $K \simeq 1.07\cdot10^{-3}$ and $\Delta n \simeq 0.028$. From this expresion, an estimate of the in-plane electrostatic field $E_x$ at the domain wall can be obtained by combining equation \ref{eq2} with the expected variation of the refractive index at a N\'{e}el DW due to the Pockels effect. Tensorial derivations for the electro-optic effect at the domain wall, accounting for the N\'{e}el rotation of the LN electro-optic tensor \cite{10.1038/ncomms15768(17)}, assuming an in-plane depolarization electrostatic field ($E_x$) oriented along the $x$-axis (i.e. orthogonal to DW) and considering only the dominant contribution of the largest electro-optic coefficient of LN ($r_{33}>>r_{13}, r_{22}$), ultimately result in the following scalar relationship between the in-built electric field and the index variation at the domain wall:
\begin{equation}
\Delta n = -\frac{1}{2}n_0^3r_{33}E_0,
\label{eq3}
\end{equation} 
where $n_0$ is the refractive index of the material in the absence of $E_x$, i.e. the ordinary refractive index of LN, and $r_{33} = 30\cdot 10^{-12}\,$m/V its electro-optic coefficient. By combining equations \ref{eq2} and \ref{eq3} we obtain an estimate of the in-plane electric field $E_x \simeq 3000\,$kV/cm, which was used as input in the calculation of the electric field by finite elements presented in the main manuscript. It is important to highlight that to reach an in-plane electric field with this strength it was necessary to introduce in the calculations a charge density of $\pm 2\,\mu$C/$\mu$m$^2$.

\subsection{Drift-diffusion continuity equations}\label{2}

The incidence of a laser beam over the 1L-MoSe$_2$ flake generates electron-hole pairs in the TMD whose kinematics is strongly influenced by the PPLN. At the center of the different domains, where $E_x$ is weak, these electron-hole pairs can easily form an exciton. However, at the DW, where $E_x$ has a large intensity, electron-hole pairs are dissociated and the resulting free carriers are efficiently separated towards opposite ferroelectric domains. In this way, the raise of the charge landscape in each domain facilitates the formation of trions and reduces the density of neutral excitons. We quantify the modulation of the particle densities across the PPLN domains with a set of five coupled differential rate equations that describe, in time, the generation, recombination, drift and diffusion of electrons, holes as well as neutral and charged excitons in a 1D toy model \cite{jimenez2012drift, Ivanov_2002}. For neutral excitons, the rate equation can be written as 
\begin{equation*}
    \frac{dn_X(x)}{dt}=G(x_0,x)[1-f(x)]+\Delta^X-\Sigma^{X^-}-\Sigma^{X^+}-\Gamma^X,
    \label{eq1}
\end{equation*}
where an exciton density $n_X(x)$ is generated via the generation rate $G(x_0,x) = Pg(x_0,x)$ of intensity $P$ and a Gaussian shape $g(x_0,x)$ at position $x_0$. This function describes the effect of the finite Gaussian beam, while the efficiency of this generation, $1-f(x)$, depends on the position on the sample. For example, directly at the boundary, this efficiency tends towards $0$, since excitons are immediately dissociated due to the large in-plane electric field \cite{Pedersen.2016,Thygesen.2017,Massicotte.2018}. The excitons diffuse away from their generation spot, described by the diffusion equation $\Delta^X = D_X\frac{d^2n_X}{dx^2}$ with the diffusion constant $D_X$. Across the sample, there is a remaining electron and hole density ($n_{e,h}$), such that an exciton can capture a charged particle and form negatively (positively) charged excitons at a rate $\Sigma^{X^-}=A\cdot n_Xn_e$($\Sigma^{X^+}=B\cdot n_Xn_h$). Finally, the neutral exciton recombines with the rate $\Gamma^X= n_X/\tau_X$, where $\tau_X$ is the exciton lifetime.

\begin{figure}
\includegraphics*[keepaspectratio=true, clip=true, angle=0, width=.6\columnwidth, trim={0mm, 0mm, 0mm, 0mm}]{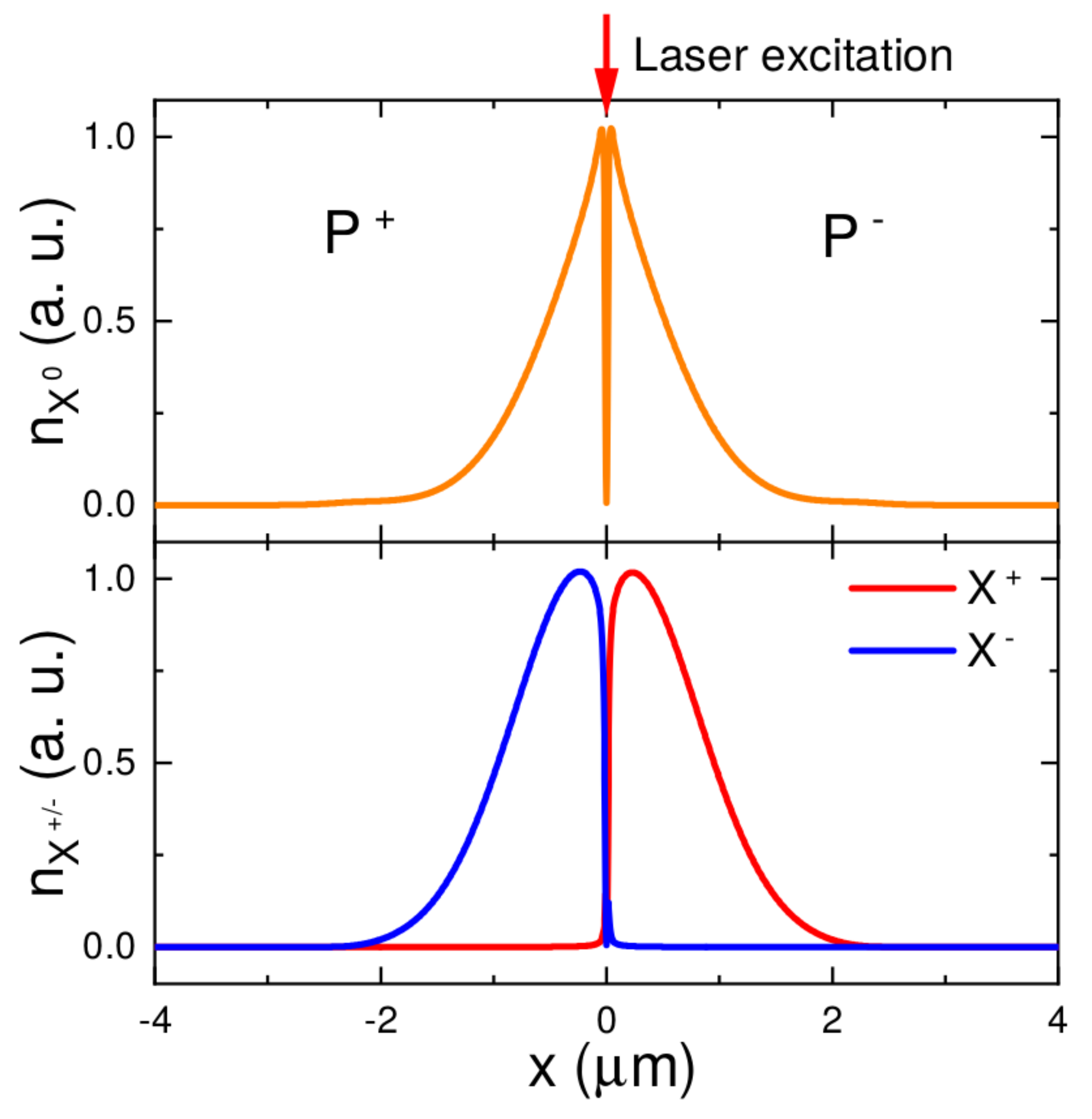}
\caption{Distribution of neutral (top) and charged (bottom) excitons generated by a Gaussian beam centered at the domain wall, $x = 0$. The spot diameter is 0.7\,$\mu$m.}
\label{figura:SM2}
\end{figure}

Since the strong in-plane field at the DW dissociates excitons into free electrons and holes, the rate equations for excess charges in the 2D material can be written as
\begin{equation*}
 \frac{dn_e(x)}{dt} = G(x_0,x)f(x)+\Delta^e+\Omega^e-\Sigma^{X^-}+\Gamma^{X^-}
    \label{eq2}
\end{equation*}
and 
\begin{equation*}
 \frac{dn_h(x)}{dt} = G(x_0,x)f(x)+\Delta^h+\Omega^h-\Sigma^{X^+}+\Gamma^{X^+},
    \label{eq3}
\end{equation*}
where the drift term for electrons and holes due to the in-plane electric field $E_x(x)$ is $\Omega^{e,h}=\mu_{e,h}[\frac{dn_{e,h}}{dx}E_x(x)+\frac{dE_x(x)}{dx}n_{e,h}]$. The recombination rate of charged excitons ($\Gamma^{X^-}$ and $\Gamma^{X^+}$) is presented in these equations as an extra source of electrons and holes, respectively. Finally, the rate equations for positively and negatively charged excitons in the system are
\begin{equation*}
\frac{dn_{X^{\pm}}(x)}{dt}=\Sigma^{X^{\pm}}+\Delta^{X^{\pm}}\pm \Omega^{X^{\pm}}-\Gamma^{X^{\pm}}. 
\label{eq4}
\end{equation*}
In the latter equation, the source of trions is the term $\Sigma^{X^{\pm}}$, equal to the rate in which an exciton captures a charge, and thus contributed negatively in the previous equations.

 Figure \ref{figura:SM2} presents $n_{X^0}$ and $n_{X^{+/-}}$ generated by a Gaussian beam $G(0,x)$, centered at the DW and calculated with the set of equations presented above. In the upper panel, the exciton density shows a  dip in its density at $x=0$ due to the huge electric field at the interface. The charged excitons are pushed out of the domain wall and are separated in different domains, as shown in the lower panel. Note that this system of equations describes the particles density distribution generated by the laser beam at a position $x_0$. Nevertheless, only the density of particles that is in the region sensed by the laser beam is observed in the PL experiments. To obtain the observed distribution of particles for a laser at $x_0$, as presented in the main text, it is necessary to spatially integrate these distributions multiplied by $g(x_0, x)$. For instance, the observed density of excitons and trions at $x=0$ is 
\begin{equation*}
n^*_{X^0}(x=0) \propto \int n_{X^0}(x) \cdot g(0,x) dx 
\label{eq5}
\end{equation*}
and
\begin{equation*}
n^*_{X^{+/-}}(x=0) \propto \int n_{X^{+/-}}(x) \cdot g(0,x) dx,
\label{eq5}
\end{equation*}
respectively.

The complete set of parameters used in the calculations is presented in table \ref{tb:par}. Those material parameters that were not found in the literature were estimated so that they correlate with those that were found in the literature. Finally, the charge capture rates ($A$ and $B$) were fitted to obtain an exciton-trion rate similar to those observed in the experiments.

\begin{table}
\caption{Parameters used in the calculations presented in the main text.}
\centering
\begin{tabular}{ c  c  c }
  \hline
  Gaussian spot diameter & $d$ & 0.7\,$\mu$m \\
  Exciton diffusion & $D_X$ & 14\,cm$^{-2}$s$^{-1}$ (\cite{PhysRevB.94.165301, kumar2014exciton}) \\
  Exciton lifetime & $\tau_X$ & 100\,ps (\cite{RevModPhys.90.021001, kumar2014exciton, PhysRevB.94.165301})\\
  Native electron doping & $n_{e0}$ & $3\times10^{10}\,$cm$^{-2}$ (\cite{10.1038/ncomms2498(13)}) \\
  Electron mobility & $\mu_e$ & 500\,cm$^2$V$^{-1}$s$^{-1}$ (\cite{shepard2017trion(17)}) \\
  Electron difussion & $D_e$ & 30\,cm$^{-2}$s$^{-1}$ \\
  Hole mobility & $\mu_h$ & 500\,cm$^2$V$^{-1}$s$^{-1}$ \\
  Hole difussion & $D_h$ & 30\,cm$^{-2}$s$^{-1}$ \\
  Negative trion mobility & $\mu_{X^-}$ & 150\,cm$^2$V$^{-1}$s$^{-1}$ \\
  Negative trion difussion & $D_{X^-}$ & 30\,cm$^{-2}$s$^{-1}$ \\
  Negative trion lifetime & $\tau_{X^-}$ & 70\,ps (\cite{PhysRevB.94.165301}) \\
  Positive trion mobility & $\mu_{X^+}$ & 150\,cm$^2$V$^{-1}$s$^{-1}$ \\
  Positive trion difussion & $D_{X^+}$ & 30\,cm$^{-2}$s$^{-1}$ \\
  Positive trion lifetime & $\tau_{X^+}$ & 70\,ps (\cite{PhysRevB.94.165301}) \\
  Positive(Negative) charge capture rate & $A(B)$ & $6\times10^{-7}\,$s$^{-1}$ \\
  \hline  
  \label{tb:par}
\end{tabular}
\end{table} 

\bibliography{Biblio}

\end{document}